
%
\input phyzzx.tex

%
\font\largemath = cmti10 scaled \magstep2
%
\REF\pw{G.Parisi and Y.Wu, Sci.Sin. {\bf 24} (1981) 483}
\REF\dh{For a review, P.H.Damgaard and H.H\"uffel, Phys.Rep. {\bf 152} (1987)
227}
\REF\noy{M.Namiki, I.Ohba and  K.Okano, Prog.Theor.Phys. {\bf 72} (1984) 350}
\REF\imk{K.Ikegami, T.Kimura and R.Mochizuki, CHIBA-EP-57
(1992)}
\REF\mohimohi{R.Mochizuki, CHIBA-EP-56 (1991)}
\REF\sa{M.Namiki and Y.Yamanaka, Prog.Theor.Phys. {\bf 69} (1983)
1764,\hfill\break
E.Gozzi, Phys.Rev. {\bf D28} (1983) 1922,\hfill\break
Y.Nakano, Prog.Theor.Phys. {\bf 69} (1983) 361,\hfill\break
C.M.Bender, F.Cooper and B.Freedman, Nucl.Phys. {\bf B219} (1983) 61}
\REF\ito{K.Ito, Proc.Imp.Acad. {\bf 20} (1944) 519}
\REF\str{R.L.Stratonovich, {\it Conditional Markov Processes and Their
Application to the Theory of Optimal Control} (Elsevier, New York, 1968)}
\REF\gra{R.Graham, Phys.Lett. {\bf 109A} (1985) 209}
\REF\moimoi{J.Zinn-Justin, {\it Quantum Field Theory and
Critical Phenomena} (Clarendon Press, Oxford,
1989),\hfill\break R.Mochizuki, Mod.Phys.Lett. {\bf A5} (1990) 2335}
\REF\mohii{R.Mochizuki, Prog.Theor.Phys. {\bf 85} (1991) 407}

\def\l{\lambda}
\def\pa{\partial}

\def\KO{[}
\def\KC{]}

\def\MCO{\Bigl\{}
\def\MCC{\Bigr\}}
\def\CO{\{}
\def\CC{\}}

%


\textfont4=\largemath
\mathchardef\lS="0453
%
%
%
\footline={\hfill-- \folio\ --\hfill}
\unnumberedchapters
\date={}
%
\pubnum={CHIBA-EP-58-REV}
\date{September 1992}
\titlepage
\title{(D+1)-Dimensional Formulation for D-Dimensional
Constrained Systems}

\author{{\it Riuji} Mochizuki }
\address{Department of Physics, Faculty of Science, \break
        Chiba University,  \break
          1-33 Yayoi-cho, Chiba 260, Japan}

\baselineskip = 12pt
\abstract{
D-dimensional constrained systems are studied with stochastic
Lagrangian and\break Hamiltonian.  It is shown that  stochastic consistency
conditions are second class constraints and  Lagrange
multiplier fields can be determined in  (D+1)-dimensional canonical
formulation.  The Langevin equations for the constrained system are
obtained as Hamilton's equations of motion where  conjugate momenta play a
part of  noise fields.        }
\def\underset#1\to#2{\mathop{#2}\limits_{#1}}

\baselineskip = 18pt

The stochastic quantization scheme was first proposed by Parisi and
Wu\refmark\pw as an alternative quantization method for gauge
theories\refmark\dh in 1981.  After a few years Namiki {\it et
al.}\refmark\noy improved the method so as to handle constrained systems and
recently more rigorous discussion was shown.\refmark\imk   In their
methods  Lagrange multiplier fields can be
determined under  stochastic consistency conditions (the consistency
conditions for the fictitious time).  If we extend the stochastic
quantization method to phase space, we can determine the Lagrange
multipliers accompanied with the first class constraints by almost the same
method.\refmark\mohimohi  They seem strange because we cannot determine
them in  canonical formalism.  In this letter we introduce the
stochastic Lagrangian\refmark\sa and Hamiltonian for constrained systems
and treat them within a (D+1)-dimensional canonical framework.  It is
shown that  stochastic consistency conditions are second class
constraints and we obtain the Langevin equations for the constrained system
as Hamilton's equations of motion.

We consider a system described by the action $S$
in D-dimensional Euclidean space-time.  In the stochastic quantization
scheme the Langevin equation is an important tool, which is written in
the discretized form as
$$
dq_i=-{\delta S\over\delta q_i}dt+dW_i.\ \ \ \ (i=1,\cdots ,N)\eqno(1)
$$
Here $t$ is the fictitious time and $dW_i$ is the Wiener process whose
expectation values are defined as
$$
\langle dW_i(t)dW_j(t^{\prime})\cdots\rangle={1\over
n}\int  D(dW)dW_i(t)dW_j(t^{\prime})\cdots\exp\MCO-\int dx\sum_m
L(dW(\tau_m))\MCC,\eqno(2)
$$
where $n$ is a normalization factor and $L(dW)$ is  stochastic
Lagrangian.  If the system has constraints
$$
F_a=0,\ \ \ \ (a=1,\cdots ,M;\
N>M)\eqno(3)
$$
we define the stochastic Lagrangian as
$$
L(dW)\equiv
{1\over 4dt}dW_idW_i+\l_adF_a,\eqno(4)
$$
where $\l_a$ is a Lagrange multiplier field.   In the stochastic
Lagrangian (4) we have changed the constraints (3) into their
differentials.  It is quite natural in this formulation because the order
of the stochastic Lagrangian is $dt$ and we rewrite it with the
Langevin equations, i.e. stochastic differential equations, as shown in the
following.

When we handle stochastic differential equations, we should make choice of
their calculation rules, that is, adopt either Ito's
calculation rule\refmark\ito or Stratonovich's one\refmark\str in a
general way.  If we adopt the latter rule, we must bother about the
Jacobian factor which appears when we change the integration variables in
the generating functional.  On the other hand the Jacobian factor is a
unity if we calculate with Ito's rule,\refmark\sa therefore we will use
it in the following calculation.  In the Ito calculus we have Ito's
formula
$$
\langle df\rangle =\langle{\pa f\over\pa q_i}dq_i\rangle+{1\over
2}\langle {\pa^2 f\over\pa q_i\pa q_j}dW_idW_j\rangle\eqno(5)
$$
where $f$ is an arbitrary functional of $q_i$'s.

To change the integration variables we insert a unity
$$
1=\int D(q)\delta\CO dq_i+{\delta S\over\delta q_i}dt-dW_i\CC
$$
into RHS of (2) and integrate over $dW$.  Then the distribution functional
$Z$ reads
$$
Z=\int D(q)\exp \MCO-\int dx\sum L(q)\MCC,\eqno(6)
$$
where
$$
L(q)={1\over 4dt}(dq_i+{\delta S\over\delta q_i}dt)^2+\l_adF_a,\eqno(7)
$$
$$
dF_a={\pa F_a\over \pa q_i}dq_i+{1\over 2}\cdot{\pa^2 F_a\over \pa
q_i\pa q_j}M_{ij}.\eqno(8)
$$
The second term in RHS of (8) is due to Ito's formula,
$M_{ij}$ in which is a $dt$-order
expectation value of $dW_idW_j$ which is not $2\delta_{ij}$ because
the Lagrange multiplier fields  should contain a noise-like
part.\refmark\imk  We will compute it later.

{}From the stochastic Lagrangian (7) we define stochastic conjugate momenta
as $$
p_i\equiv{\pa L\over\pa(dq_i)}={1\over 2}({dq_i\over dt}+{\pa L\over \pa
q_i})+\l_a{\pa F_a\over\pa q_i},\eqno(9.a)
$$
$$
\pi_a\equiv{\pa L\over\pa(d\l_a)}\approx 0.\eqno(9.b)
$$
The latter equation is a primary constraint, and hence the stochastic
Hamiltonian $H$ is
$$
\eqalign{
H\equiv & p_idq_i+\pi_ad\lambda_a-L+u_a\pi_adt\cr
=& p_i(2p_idt-{\delta S\over \delta q_i}dt-2\l_a{\pa F_a\over\pa
q_i}dt)-{1\over dt}(p_idt-\l_a{\pa F_a\over\pa q_i}dt)(p_idt-\l_b{\pa
F_b\over\pa q_i}dt)\cr &-\l_a\MCO{\pa F_a\over\pa q_i}(2p_idt-{\delta
S\over \delta q_i}dt-2\l_b{\pa F_b\over\pa q_i}dt)+{1\over 2}\cdot{\pa^2
F_a\over\pa q_i\pa q_j}M_{ij}\MCC +u_a\pi_adt,\cr
}\eqno(10)
$$
where $u_a$ is the other Lagrange multiplier into which $d\l_a/dt$ is
absorbed.

The stochastic differentials are defined using Poisson's bracket
$$
\CO A,B\CC\equiv{\pa A\over\pa q_i}{\pa B\over\pa p_i}-{\pa
A\over\pa p_i}{\pa B\over\pa q_i}+{\pa A\over\pa\l_a}{\pa B\over\pa
\pi_a}-{\pa A\over\pa \pi_a}{\pa B\over\pa\l_a}
$$
as
$$
dA\equiv\CO A,H\CC.\eqno(11)
$$
The stochastic consistency condition\refmark\noy of $\pi_a\approx
0$ thus reads
$$
\eqalign{
d\pi_a&=\CO\pi_a,H\CC\cr
&={\pa F_a\over \pa q_i}(2p_idt-{\delta S\over\delta q_i}dt)+{1\over
2}\cdot{\pa^2 F_a\over\pa q_i\pa q_j}M_{ij}-2{\pa F_a\over\pa q_i}{\pa
F_b\over\pa q_i}\l_bdt\cr
&\equiv\phi_a\approx 0.\cr}\eqno(12)
$$
This is a secondary constraint.  If we determine $\l_b$ so as to satisfy
it, we need no more constraints.  We may identify $\phi_a$ with $dF_a$
and $\pi_a$ with $F_a+C_a$ where $C_a$ is an arbitrary functional
independent of the fictitious time.  Consequently we should choose the
initial conditions as $C_a=0$ to realize the constraints (3).  The Poisson's
bracket of the two constraints (9.b) and (12) is  $$
\CO\pi_a,\phi_b\CC=2{\pa F_a\over\pa q_i}{\pa F_b\over\pa q_i}\equiv
2D_{ab}.\eqno(13)  $$
Here we assume that $D_{ab}$ is not zero and has its
inverse $D^{-1}_{ab}$.\refmark\noy  Then the above equation shows that
both of them are second class constraints.  When we write the total
Hamiltonian $H_T$ as
$$
H_T=H+v_a\pi_adt+w_a\phi_a\eqno(14)
$$
where $v_a$ and $w_a$ are Lagrange multiplier fields, we can determine
them with the help of the stochastic consistency conditions again.  They read
$$
w_a=0,
$$
$$
v_a={1\over 2}D^{-1}_{ab}\CO\phi_b,H\CC.
$$
Accordingly the total Hamiltonian becomes
$$
H_T=H+{1\over 2}D^{-1}_{ab}\CO\phi_b,H\CC\pi_a.\eqno(15)
$$

Next we derive Hamilton's equation of motion for $q_i$.  If we use
Dirac's bracket
$$
\KO A,B\KC\equiv\CO A,B\CC-\CO
A,\Phi_r\CC\Delta^{-1}_{rs}\CO\Phi_s,B\CC\eqno(16)
$$
with
$$
\Phi_r\equiv\pi_a,\ \phi_a,
$$
$$
\Delta_{rs}\equiv\CO\Phi_r,\Phi_s\CC,
$$
we may replace the weak equality $\approx$
by the strong equality $=$.  In the following we always use Dirac's
bracket instead of Poisson's bracket and rewrite the total Hamiltonian
by the   aid of the second class constraints (9.b) and (12) as
$$
\eqalign{
H_T=p_ip_idt-& p_i{\delta S\over\delta q_i}dt\cr
-{1\over 4dt}D^{-1}_{ab}&\MCO{\pa F_a\over \pa
q_i}(2p_idt-{\delta S\over\delta q_i}dt)+{1\over 2}\cdot{\pa^2 F_a\over\pa
q_i\pa q_k}M_{ik}\MCC\cr \times &\MCO{\pa F_b\over \pa
q_j}(2p_jdt-{\delta S\over\delta q_j}dt)+{1\over 2}\cdot{\pa^2 F_b\over\pa
q_j\pa q_l}M_{jl}\MCC .\cr}\eqno(17) $$
Using it we obtain Hamilton's equation of motion for $q_i$:
$$
\eqalign{
dq_i&=\KO q_i,H_T\KC\cr
&=-K_{ij}{\delta S\over\delta q_j}dt+2K_{ij}p_jdt-{1\over 2}\cdot{\pa
F_a\over\pa
q_i}D^{-1}_{ab}{\pa^2 F_b\over\pa q_j\pa q_k}M_{jk}\cr}\eqno(18)
$$
with
$$
K_{ij}\equiv\delta_{ij}-{\pa F_a\over\pa
q_i}D^{-1}_{ab}{\pa F_b\over\pa q_j},
$$
which of course satisfies $dF_a=0$, and hence $(N-M)$ independent
equations exist.

The equation (18) is  Langevin equation if we may identify
$2K_{ij}p_jdt$ with  noise fields.  In the following we show that is the
case.  In order to decompose the variables into constraint variables
and independent ones,\refmark\noy\refmark\imk we introduce a new set of
variables $\CO Q^{\mu}\CC$, ($\mu =
1,2,\cdots ,N$), which are defined as
$$
{\partial Q^{\mu}\over\partial q_i}=e^{\mu}_{\ i},\eqno(19)
$$
where $e^{\mu\ \prime}_{\ i}$s are vielbein fields defined as
$$
e^a_{\ i}\equiv{\partial F_a\over\partial q_i}\ ,\ \ \ \ \ \ e_{a,i}\equiv
D^{-1}_{ab}{\partial F_b\over\partial q_i},\ \ \ \ (a=N-M+1,\cdots
,N)\eqno(20.a)
$$
$$
e^A_{\ i}e_{a,i}=0, \ \ \ \ \ \ \ \ \ \ \ \ \ \ \ \ (A=1,2,\cdots
,N-M)\eqno(20.b)
$$
$$
e_{A,i}\equiv (g^{-1})_{AB}e^B_{\ i},\ \ \ \ \ g^{AB}\equiv e^A_{\ i}e^B_{\ i}.
\eqno(20.c)
$$
Here we have assumed that the metric $g^{AB}$ is non-singular.  These
definitions lead to following relations
$$
e^{\mu}_{\ i}e_{\nu ,i}=\delta_{\mu\nu},\ \ \ \ \ e^{\mu}_{\ i}e_{\mu ,j}
=\delta_{ij},\eqno(21.a)
$$
$$
e^A_{\ i}e^a_{\ i}=e^A_{\ i}e_{a, i}=e^{A,i}e^a_{\
i}k=e_{A,i}e_{a,i}=0,\eqno(21.b)
$$
$$
e^A_{\ i}e_{A,j}=K_{ij},\ \ \ \ \ \ \ e^a_{\
i}e_{a,j}=\delta_{ij}-K_{ij},\eqno(21.c)
$$
$$
K_{ij}e^a_{\ j}=K_{ij}e_{a,j}=0,\ \ (\delta_{ij}-K_{ij})e^A_{\
j}=(\delta_{ij}-K_{ij})e_{A,j}=0,\eqno(21.d)
$$
$$
det(e^{\mu}_{\ i})\not= 0.\eqno(21.e)
$$
The equations (21.e) shows that a manifold spanned by $Q^{\mu}$'s is equivalent
to one
by $q_i$'s.

Assuming $2K_{ij}p_jdt$ to be the noise field and to give the same expectation
value
 as $dW_i$, which assumption will be confirmed later, we can derive the
Langevin equations
 for the new variables.  Taking account of Ito's formula (5), they
 are
$$
dQ^{\mu}=e^{\mu}_{\ i}dq_i+{1\over 2}{\partial e^{\mu}_{\ i}\over\partial q_j}
M_{ij}.\eqno(22)
$$
For the constraint variable $Q^a$=$F_a$, we have
$$
dQ^a=0\eqno(23.a)
$$
and the initial conditions should be chosen to be zero to realize the
constraints (3),.
  On the other hand, for the independent variables $Q^{A\prime}$s, the
Langevin equations are
$$
dQ^A=-g^{AB}{\delta S\over\delta Q^B}dt+e^A_{\ i}(2p_idt)+{1\over 2}{\partial
 e^A_{\ i}\over\partial Q^B}e^B_{\ j}M_{ij}.\eqno(23.b)
$$
Moreover if we decompose the vielbein $e^A_{\ i}$ as
$$
e^A_{\ i}\equiv E^A_{\ I}\epsilon^I_{\ i},\ \ \ \ \ (I=1,2,\cdots N-M)
\eqno(24.a)
$$
$$
\epsilon^I_{\ i}\epsilon^J_{\ i}=\delta^{IJ},\ \ \ \ e^A_{\ i}e^B_{\ i}=
E^A_{\ I}E^B_{\ J}\delta^{IJ}=g^{AB},\eqno(24.b)
$$
where $\epsilon^I_{\ i}$ only rotates the tangent
space spanned by the independent variables,
we may change the second term in RHS of the Langevin equation (23.b) into
fields with $N-M$ indices:
$$
e^A_{\ i}(2p_idt)=E^A_{\ I}d\Omega^I,\eqno(25)
$$
where $d\Omega^I$ is defined as
$$
d\Omega^I\equiv\epsilon^I_{\ i}(2p_idt),\eqno(26)
$$
which will be regarded as the Wiener process as shown in the following.

We change the variables $q_i$'s in the distribution functional (6)
 into the new ones $Q^{\mu}$'s
$$
Z=\int D(Q)det(e_{\mu,i})\exp\MCO -\int dx\sum L(Q)\MCC,\eqno(27)
$$
and substitute zero for $Q^a$'s.
Then again we insert a unity
$$
1=\int D(d\Omega)det(E_{\ I}^A)\delta\CO
dQ^A+g^{AB}{\delta S\over\delta Q^B}dt-E^A_{\ I}d\Omega^I-{1\over 2}{\partial
 e^A_{\ i}\over\partial Q^B}e^B_{\ j}M_{ij}\CC
$$
into the distribution  functional (27) and integrate over
$Q^A$'s.  Neglecting the normalization
factor, (27) becomes
$$
Z=\int D(d\Omega)\exp\MCO -\int dx\sum L(d\Omega)\MCC,\eqno(28)
$$
where
$$
L(d\Omega)={1\over 4dt}d\Omega_Id\Omega_I.\eqno(29)
$$
The expectation values are easily computed with the above distribution
functional:
$$
\langle d\Omega^I\rangle =0,
$$
$$
\langle d\Omega^I d\Omega^J\rangle=2\delta^{IJ}dt,
$$
and
$$
M_{ij}=2\langle e^A_{\ i}e_{A,j}\rangle dt.
$$
The Langevin equation (23.b) thus becomes
$$
dQ^A=-g^{AB}{\delta S\over\delta Q^B}dt+E^A_{\ I}d\Omega^I+{\partial
 e^A_{\ i}\over\partial Q^B}e^B_{\ i}dt,\eqno(30)
$$
which agrees with the Langevin equation for constrained systems obtained
by  the other method in the stochastic quantization scheme.\refmark\imk  The
last term is necessary for the general coordinate transformation
covariance\refmark\gra and, in perturbation
theories, cancellation of some of divergent terms.\refmark\imk  The proper
Fokker-Planck equation on ($N-M$)-dimensional constraint surface is given
by the ordinary prescription with the Langevin equations (23).  It will be
straightforward to extend the results to the phase-space stochastic
quantization scheme.\refmark\mohimohi

Finally we comment on the results when we obey Stratonovich's calculation
rule.  In this case the Langevin equation (30) should be a little
modified.\refmark\imk\refmark\moimoi\refmark\mohii  Nevertheless it, of course,
does not mean the two
rules lead different physical results.  For example the Fokker-Planck
equation, whose equilibrium distribution should be consistent with the
path-integral form, is the same in both the cases.

\section{Acknowledgments}
I would like to thank Professor T.Kimura, Dr. A.Nakamura and K.Ikegami for
useful discussions.

\vfill\eject

\refout

\end